\documentclass{article}

\usepackage{arxiv}

\usepackage[main=english,ngerman]{babel}
\usepackage[utf8]{inputenc} 
\usepackage[T1]{fontenc}    
\usepackage{hyperref}       
\usepackage{url}            
\usepackage{booktabs}       
\usepackage{amsfonts}       
\usepackage{nicefrac}       
\usepackage{microtype}      
\usepackage{lipsum}		

\usepackage{tikz}
\usetikzlibrary{decorations.pathreplacing}

\title{Query-Sequence Optimization on a Reconfigurable Hardware-Accelerated System}

\author{
  Lekshmi B.G., Andreas Becher, Klaus Meyer-Wegener \\
    Friedrich-Alexander-Universität Erlangen-Nürnberg (FAU)\\
 91058, Erlangen, Germany \\
  \texttt{\{lekshmi.bg.nair, andreas.becher, klaus.meyer-wegener\} @fau.de} \\
}

\begin{document}
\maketitle

\begin{abstract}
Hardware acceleration of database query processing
can be done with the help of FPGAs.
In particular, they are partially reconfigurable during runtime,
which allows for the runtime adaption of the hardware to a variety of queries.
Reconfiguration itself, however, takes some time.
As the affected area of the FPGA is not available for computations during the reconfiguration,
avoiding some of the reconfigurations can improve overall performance.
This paper presents optimizations based on query sequences,
which reduces the impact of the reconfigurations.
Knowledge of coming queries is used to
(I) speculatively start reconfiguration already when a query is still running and
(II) avoid overwriting of reconfigurable regions that will be used again in subsequent queries.
We evaluate our optimizations with a calibrated model and measurements for various parameter values.
Improvements in execution time of up to 21\% can be obtained
even with sequences of only two queries.
\end{abstract}


\maketitle

\section{Introduction}

\label{sec:introduction}

There are already a couple of projects addressing
the acceleration of database query processing
with the help of FPGAs, e.\,g.\ \cite{Muel09a,Sukh13a,Owai19a},
and their integration into a DBMS.
The ReProVide project \cite{Bech19a} is one of them.
The particular approach of this project is
to use dynamic reconfiguration of the FPGA
in a combination of
a novel DBMS optimizer and
accelerated data storage units called ReProVide Processing Units (RPUs).
On these RPUs,
a library of query-processing modules is available,
which can be configured onto the FPGA in the order of 15ms.
Due to the limited amount of logic resources on an FPGA,
not all modules can be made available simultaneously.
So at some point in time,
only a subset is ready for use in query processing.
Hence, reconfiguration of one or more region of the FPGA is needed
to process an incoming query optimally,
if the required modules are not loaded already.

The idea presented in this paper is
that when a sequence of queries to be executed repeatedly is known,
information about this sequence can be given to the RPU
via so-called \emph{hints}.
The RPU can use this information
to reduce the overall execution time of the sequence of queries.

In the following,
we will indicate where such sequences of queries may come from.
They are assumed to be part of an application program
that e.\,g. fills the frames of some modular screen output
from different parts of an underlying database.

\section{The ReProVide Processing Unit (RPU)}
\label{sec:reprovide}

This section gives some more detail on
how the ReProVide system does the query processing.
It is a ``system on chip'' (SoC)
with own storage (SSD),
an ARM processor,
and some memory (DRAM)---in addition to the FPGA.
So the RPU ``owns'' data tables,
which means that any access to these tables
goes through the RPU\footnote{In this paper, we neglect the caching of tables in a host system.}.
The RPU is attached
to a host running a (relational) DBMS
via a fast network.
It is important to note that
streaming the table data through the FPGA
comes at no additional cost\footnote{We assume $datarate_{network} \leq datarate_{storage}$}.

However, reconfiguration may be required
before the RPU can process the data in the way
a particular query demands.
The FPGA of a RPU contains various static hardware modules,
like a storage controller, a network controller, data interconnects, and local memories,
as well as multiple partially reconfigurable regions (PRs) \cite{Bech19a}.
Data is processed by so-called accelerators loaded into these PRs.
RPUs execute a partial query
by streaming the tables from the storage
at line-rate
through one or multiple accelerators
to the network interface.
Operations like sorting or joining of larger tables
cannot be implemented in an efficient way on such a streaming architecture
and are therefore left to the DBMS.
As the accelerators are optimized for line-rate processing,
and FPGA resources are limited,
not all available operator modules can be combined into a single accelerator.
E.\,g.\ implementations of arithmetic operators differ
not only in the operation they implement (mult, add, \ldots ),
but also in the type they operate on (float, int32, int64).
Dynamic partial reconfiguration allows
to offer support for more and more operators
with a growing library of available accelerators,
each implementing a reasonable subset of all available operators.

This means the RPU has a state,
which consists of the set of accelerators configured,
and a cost of changing that state,
introduced by the time it takes to exchange accelerators
through the reconfiguration of PRs.

A special approach of the ReProVide project is
that the interface of the RPU will allow to send some \emph{hints}
in addition to the query-execution request.
These hints do not change any functionality,
but give some information to the RPU
so it can optimize the execution further.
This paper introduces one such hint
to avoid unnecessary reconfiguration
or to begin with reconfiguration for coming queries
while still executing the current one.

The optimizer of the DBMS that hosts the RPU
identifies the filter operations that can be pushed down to the RPU.
Since the data has to be retrieved from the table storage anyway,
and since the processing is almost for free,
all operators that can be executed by the RPU
are pushed down.
This may greatly reduce the amount of transferred data,
and thus not only reliefs the network of unneeded traffic,
but also reduces the load on the DBMS,
as less data needs to be processed by it.

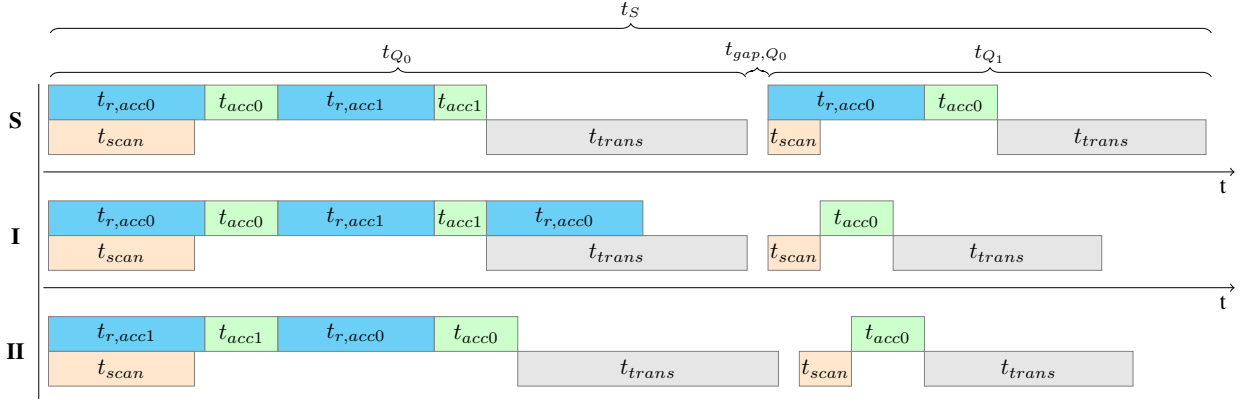
\begin{figure*}[t]
	\begin{center}
		\resizebox{\textwidth}{!}{
\newcommand{\job}[4][fill=black]{

    \def\tscale{0.45}

    \foreach \x in {#4}
    {
        \pgfmathparse{(\x)*\tscale}
        \node[rectangle,draw=gray,#1,minimum height=1.5em,minimum width=#3*1em*\tscale,inner sep=0,anchor=west] (j\x#2) at (\pgfmathresult,#2) {};
    }
}%
\newcommand{\drawhelper}[3][gray]{
    \path (#2);
    \pgfgetlastxy{\startX}{\startY};
    \path (#3);
    \pgfgetlastxy{\endX}{\endY};

    \pgfmathsetmacro\startX{int(\startX*1pt/1em}

    \pgfmathsetmacro\endX{int(\endX*1pt/1em)}

    \foreach \x in {\startX,...,\endX}
    {
        \draw[draw=#1] (\x,\startY) -- (\x,\endY) node[below] {\scriptsize \x};
    }
}%
\newcommand{\advancet}[1][\t]{
\pgfmathsetmacro\pos{\pos+#1}
}%
\newcommand{\savepos}[1][\prevpos]{
\pgfmathsetmacro#1{\pos}
}%
\begin{tikzpicture}[y=1em,x=1em]


                \colorlet{reconfcolor}{cyan!50!white}
                \colorlet{iocolor}{orange!20!white}
                \colorlet{execcolor}{green!20!white}
                \colorlet{netcolor}{gray!20!white}

                \pgfmathsetmacro\pos{2}
                \def\y{0}
                \def\t{15}
                \job[draw,fill=reconfcolor,label={center:$t_{r,acc0}$}]{\y}{\t}{\pos}
                \savepos
                \advancet
                \def\t{14}
                \job[draw,fill=    iocolor,label={center:$t_{scan}$}]{\y-1.5}{\t}{\prevpos}
                \savepos

                \def\t{7}
                \job[draw,fill=  execcolor,label={center:$t_{acc0}$}]{\y-0.0}{\t}{\pos}
                \advancet
                \def\t{15}
                \job[draw,fill=reconfcolor,label={center:$t_{r,acc1}$}]{\y}{\t}{\pos}
                \advancet
                \def\t{5}
                \job[draw,fill=  execcolor,label={center:$t_{acc1}$}]{\y-0.0}{\t}{\pos}
                \advancet

                \def\t{25}
                \job[draw,fill=netcolor,label={center:$t_{trans}$}]{\y-1.5}{\t}{\pos}
                \advancet
                \advancet[2]

                \savepos
                \def\t{15}
                \job[draw,fill=reconfcolor,label={center:$t_{r,acc0}$}]{\y}{\t}{\pos} 
                \advancet
                
                \def\t{5}
                \job[draw,fill=    iocolor,label={center:$t_{scan}$}]{\y-1.5}{\t}{\prevpos}
                
                \def\t{7}
                \job[draw,fill=  execcolor,label={center:$t_{acc0}$}]{\y-0.0}{\t}{\pos}
                \advancet
                \def\t{20}
                \job[draw,fill=netcolor,label={center:$t_{trans}$}]{\y-1.5}{\t}{\pos} 

                \draw[->] (0.7,\y-3) -- (52.1,\y-3) node[below left] {t};

                \pgfmathsetmacro\pos{2}
                \def\y{-5}
                \def\t{15}
                \job[draw,fill=reconfcolor,label={center:$t_{r,acc0}$}]{\y}{\t}{\pos}
                \savepos
                \advancet
                \def\t{14}
                \job[draw,fill=    iocolor,label={center:$t_{scan}$}]{\y-1.5}{\t}{\prevpos}
                \savepos

                \def\t{7}
                \job[draw,fill=  execcolor,label={center:$t_{acc0}$}]{\y-0.0}{\t}{\pos}
                \advancet
                \def\t{15}
                \job[draw,fill=reconfcolor,label={center:$t_{r,acc1}$}]{\y}{\t}{\pos}
                \advancet
                \def\t{5}
                \job[draw,fill=  execcolor,label={center:$t_{acc1}$}]{\y-0.0}{\t}{\pos}
                \advancet
                \savepos[\savedpos]

                \def\t{25}
                \job[draw,fill=netcolor,label={center:$t_{trans}$}]{\y-1.5}{\t}{\pos}
                \advancet

                \advancet[2]
                \def\t{15}

                \job[draw,fill=reconfcolor,label={center:$t_{r,acc0}$}]{\y-0}{\t}{\savedpos} 
                \def\t{5}
                \job[draw,fill=    iocolor,label={center:$t_{scan}$}]{\y-1.5}{\t}{\pos} 
                \advancet
                \def\t{7}
                \job[draw,fill=  execcolor,label={center:$t_{acc0}$}]{\y}{\t}{\pos} 
                \advancet
                \def\t{20}
                \job[draw,fill=netcolor,label={center:$t_{trans}$}]{\y-1.5}{\t}{\pos} 

                \draw[->] (0.7,\y-3) -- (52.1,\y-3) node[below left] {t};
                
                \pgfmathsetmacro\pos{2}
                \def\y{-10}
                \def\t{15}
                \job[draw,fill=reconfcolor,label={center:$t_{r,acc1}$}]{\y}{\t}{\pos}
                \savepos
                \advancet
                \def\t{14}
                \job[draw,fill=    iocolor,label={center:$t_{scan}$}]{\y-1.5}{\t}{\prevpos}
                \savepos

                \def\t{7}
                \job[draw,fill=  execcolor,label={center:$t_{acc1}$}]{\y-0.0}{\t}{\pos}
                \advancet
                \def\t{15}
                \job[draw,fill=reconfcolor,label={center:$t_{r,acc0}$}]{\y}{\t}{\pos}
                \advancet
                \def\t{8}
                \job[draw,fill=  execcolor,label={center:$t_{acc0}$}]{\y-0.0}{\t}{\pos}
                \advancet
                \savepos[\savedpos]

                \def\t{25}
                \job[draw,fill=netcolor,label={center:$t_{trans}$}]{\y-1.5}{\t}{\pos}
                \advancet

                \advancet[2]
                \def\t{15}

                \def\t{5}
                \job[draw,fill=    iocolor,label={center:$t_{scan}$}]{\y-1.5}{\t}{\pos} 
                \advancet
                \def\t{7}
                \job[draw,fill=  execcolor,label={center:$t_{acc0}$}]{\y}{\t}{\pos} 
                \advancet
                \def\t{20}
                \job[draw,fill=netcolor,label={center:$t_{trans}$}]{\y-1.5}{\t}{\pos} 


                \draw[] (0.5,0.8) -- (0.5,-12.8) ;

                \draw [decorate,decoration={brace,amplitude=04pt},xshift=0.0em,yshift=1pt]
                (1.0,1.0) -- (31.0,1.0) node [black,midway,yshift=1.0em] 
                {\footnotesize $t_{Q_0}$};

                \draw [decorate,decoration={brace,amplitude=03pt},xshift=0.0em,yshift=2pt]
                (31.0,1.0) -- (32.0,1.0) node [black,midway,yshift=1.0em] 
                {\footnotesize $t_{gap,Q_0}$};

                \draw [decorate,decoration={brace,amplitude=04pt},xshift=0.0em,yshift=1pt]
                (32.0,1.0) -- (51.1,1.0) node [black,midway,yshift=1.0em] 
                {\footnotesize $t_{Q_1}$};

                \draw [decorate,decoration={brace,amplitude=04pt},xshift=0.0em,yshift=2em]
                (1.0,1.0) -- (51.1,1.0) node [black,midway,yshift=1.0em] 
                {\footnotesize $t_S$};

                \node[] at (-0.5, -0.75) {\bf S};
                \node[] at (-0.5, -5.75) {\bf I};
                \node[] at (-0.5,-10.75) {\bf II};

\end{tikzpicture}%
}
	\end{center}
	\caption{Sequence diagram of an example query sequence $S$ on top
		and our proposed optimizations for this sequence below,
		namely (I) speculative reconfiguration and (II) accelerator reordering.
		\label{fig:opts}}
\end{figure*}

\section{Related Work}
\label{sec:related-work}

The Hybrid Query Processing Engine (HyPE) \cite{Bres12b,Bres13a,Bres12a}
is a self-tuning optimizer framework.
As the name indicates, it allows for hybrid query processing,
that is, utilizing multiple processing devices.
CoGaDB has used it to build a hardware-oblivious optimizer,
which learns cost models for database operators
and efficiently distributes a workload to available processors.
It is hardware- and algorithm-oblivious,
i.\,e.\ it has only minimal knowledge of the underlying processors or implementation details of operators.
To achieve this,
HyPE has three components:
estimation component, algorithm selector, and hybrid query optimizer.
Simple regression models are used to estimate the behavior of hardware
w.\,r.\,t.\ properties of input data like size and selectivity.
Runtime monitoring allows verifying the predictions.
The optimizer must select an algorithm
(including the processor), estimate the cost, and decide on a plan according to heuristics.
In contrast to this work, HyPE does not address multi-query optimization and hints.

The ADAMANT project \cite{Bech18b}
enhances a DBMS
with extensible, adaptable support for heterogeneous hardware.
The main objectives are
to find a useful abstraction level
that considers the fundamental differences between the heterogeneous accelerators,
the development of an improved optimization process
that copes with the explosion of search space
with new dimensions of parallelism, and
to fine-tune device-dependent implementation parameters
to exploit the different features available in heterogeneous co-processors.
The project mainly uses OpenCL and emphasizes GPUs, but is also beginning to look into FPGAs.

Multi-query optimization
has been studied for some time already \cite{Chau16a,Chen98a,Sell88a}.
In all cases, however, the queries to be optimized are available at the same time.
Here, we receive the queries one after the other, with some time gap in between.
The template of the coming queries is known,
but the parameters, e.\,g.\ the constants used in comparisons, are not.
So we can use the techniques for identifying common subexpressions,
but we do not consider to change the ordering of the queries in this work.

Query processing using FPGA-based hardware accelerators
has been studied extensively
since a few years \cite{Naja13a,Muel09a,Bech15a}.
Offloading query operations
for reducing energy and for fast execution
is well investigated \cite{Wood14a,Zien16a,Bech16a}.
But the works are all focusing on the execution of a single query.
Here we are interested in the execution of a sequence of queries.

\section{Query-Sequence Model}
\label{sec:model}

The information base that drives the optimization is
a sequence of (relational) queries to be executed repeatedly.
It can be obtained from database query logs \cite{Wahl16a,Schw16a,Wahl18a}
or by code analysis \cite{Smit11a,Nagy13a},
including the frequency and the time of arrival.
The sequence can be adjusted manually
to focus on the most important queries,
if necessary.

We denote a query sequence $S$ containing $n$ queries
as an ordered set of individual queries $Q_0, \dots, Q_{n-1}$.
All queries are run to completion before the next starts
without any overlap in their execution.
The time between two successive queries
is estimated from experience
or is learned by repeated observations.
For any query sequence $S$,
the time gap $t_{gap,q}$ with $q \in \{Q_0, \dots, Q_{n-2}\}$ gives
the average time between two consecutive queries,
namely $q$ and its successor.
Please note that the optimization target in this model
is not to optimize the execution time of a single query,
but the time of the whole sequence,
that is, the time from the arrival of $Q_0$
until the transmission of the last result of $Q_{n-1}$,
including all time gaps.

The analysis of the queries in the sequence then
determines
\begin{itemize}
	\item the tables and attributes accessed, and in which order, as well as 
	\item the operators they request on the attributes of the tables
	(arithmetic and/or comparisons)
\end{itemize}
While the proposed approach is suitable for arbitrary operators,
we focus on filters (selections) together with arithmetic for now.
They can easily be extracted
from query-execution plans (QEPs).
The constants used in the arithmetic or in the comparisons
may come from program variables,
so they will be different in future executions of the query sequence.
We will introduce parameters here,
as known from prepared queries and stored procedures.

The query sequences are generated by application programs
that e.\,g. compose output from various parts of a database.
Imagine an e-mail client displaying overview lists
together with a preview of the first e-mail on the list
and maybe some attachments.
The first query retrieves the list elements with selected information
such as sender and subject line.
Using the id of the top line,
the second query fetches more details,
e.\,g.\ the list of other recipients and the first 10 lines of the body.
The third query finally obtains previews of the attachments.
One can easily imagine other client interfaces
consisting of frames
filled with related contents
extracted by queries from a database.

\section{Execution Model}
\label{sec:exec_model}

In order to keep the model simple,
we only use a single partial region within the RPU and do not overlap table scanning with accelerator execution.
We, however allow the parallel \emph{scan}ning of tables and reconfiguration of the partial region.
The upper part of Fig~\ref{fig:opts} depicts the execution of a query sequence $S$ with two consecutive Queries ($Q_0,Q_1$).
The execution time of a Query $t_{q}$ is the sum of the needed accelerator runtimes ($t_{accX}$) and network transport time ($t_{trans}$).
As the table scan can be executed while the first accelerator is reconfigured ($t_{r,accX}$), only the maximum value of $t_{scan}$ and $t_{r,accX}$ is added to the execution time of
its particular Query.

\begin{figure}
	\begin{center}
   \includegraphics[width=\linewidth, height=15em]{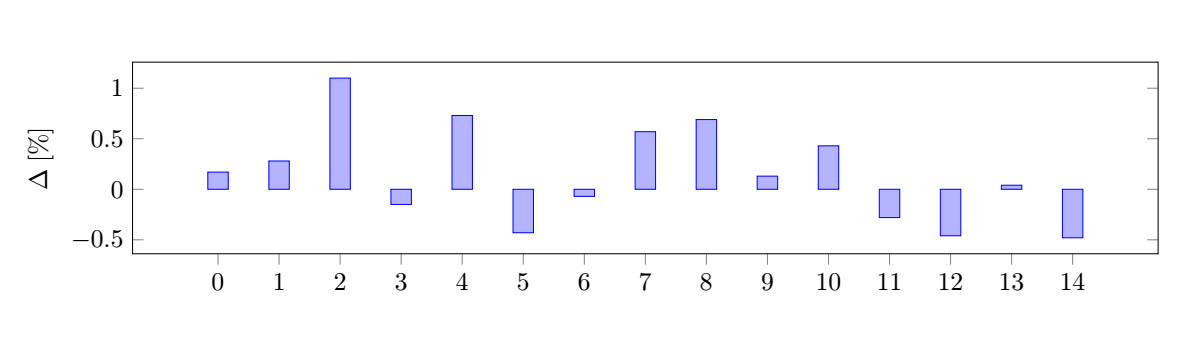}
	\end{center}	
	\caption{
		Delta of predicted execution time by emulation and RPU execution time for 15 different testqueries
		\label{fig:emulatorBehavior}
	}
\end{figure}

Build on top of cost models from Ziener~Etal. \cite{Zien16a}, a software emulator was build to test the behavior of different optimizations strategies. 
We used 15 example Queries to evaluate the correctness of the implemented emulator.
The Queries have varying selectivity from 0\% up to 100\%, 
access tables with a size factor from $1$ to $6$ in cardinality,
selection and where clauses.
Fig.\ref{fig:emulatorBehavior} shows the percentage of difference in predicted execution time by emulator and the RPU.

\section{Optimization}
\label{sec:optimization}

The optimizer of the DBMS
analyzes the sequence of pushed-down operations for similarity,
that is, common subexpressions \cite{Chen98a}.
Since the goal at this point is to avoid unnecessary reconfigurations,
only information about subsequences of common comparisons is passed to the RPU as hints.
This information can be organized
with the help of the query repository \cite{Schw16a}.

There is no scheduling of queries;
a query sequence may begin at any time.
The DBMS optimizer tries to recognize the first query of a sequence
and then retrieves the relevant sequence information,
i.\,e.\ the similarities in the sequence of pushed-down comparisons.

From that, the optimizer generates the hints to the RPU
in order to avoid reconfiguration
and thus reduce the execution time
of the pushed-down query plans.
The RPU should then
\begin{itemize}
	\item[\textbf{I}] speculatively load accelerators for subsequent queries,
	once other accelerators are no longer used by the current query, and
	\item[\textbf{II}] avoid the replacement of reusable accelerators.
\end{itemize}

The effect of optimization I is shown in the middle of Figure~\ref{fig:opts}.
While the processing of the first query of the sequence ($Q_0$) has not changed,
the reconfiguration back to Accelerator 0 ($acc0$)
is speculatively started as soon as Accelerator 1 ($acc1$) has finished.
In this example,
the reconfiguration---which runs in parallel to the result transmission---has
completed before the subsequent query ($Q_1$) arrives.
This is the optimal case for this optimization.
Query $Q_1$ can start immediately when it arrives,
without any waiting time introduced by reconfiguration.

Optimization II tries to avoid the third reconfiguration at all
by swapping the accelerator invocations of $Q_0$
if possible.
For instance, two accelerators implementing filters
can be swapped safely.
The local optimizer would invoke the filter with the lowest selectivity first
to reduce the data volume early.
This may be revoked by the swapping.
In the lower part of Figure~\ref{fig:opts} one can see the consequences.
Query $Q_0$ now needs more time than without optimization II,
as the second accelerator $acc0$ takes longer to process the data produced by $acc1$.
While this looks detrimental to the goal of optimization,
the third reconfiguration has in fact vanished
and thus the execution time for the complete sequence is reduced.

Of course this only makes sense,
if the RPU actually reacts to the hints,
that is, if the optimization works on the RPU side as well,
because it can skip some time-consuming action
or exploit waiting times.


\section{Evaluation}
\label{sec:evaluation}

We have evaluated our proposed optimizations
using the calibrated emulator from Section~\ref{sec:exec_model}.
A parameterized version of the query sequence $S$
as presented in Fig.~\ref{fig:opts}
is used for the evaluation.
It is a minimum-size sequence
and hence, any other sequence with more accelerator runs or queries
will only enhance the possibility to apply the optimizations presented here.
We vary the size of the relation to be processed by the two queries (scale factor)
and the time gap between the queries ($t_{gap,Q_0}$).

\begin{figure}[htpb]
	\begin{center}
%
%
%
%
%
 \includegraphics[width=\linewidth, height=.35\columnwidth]{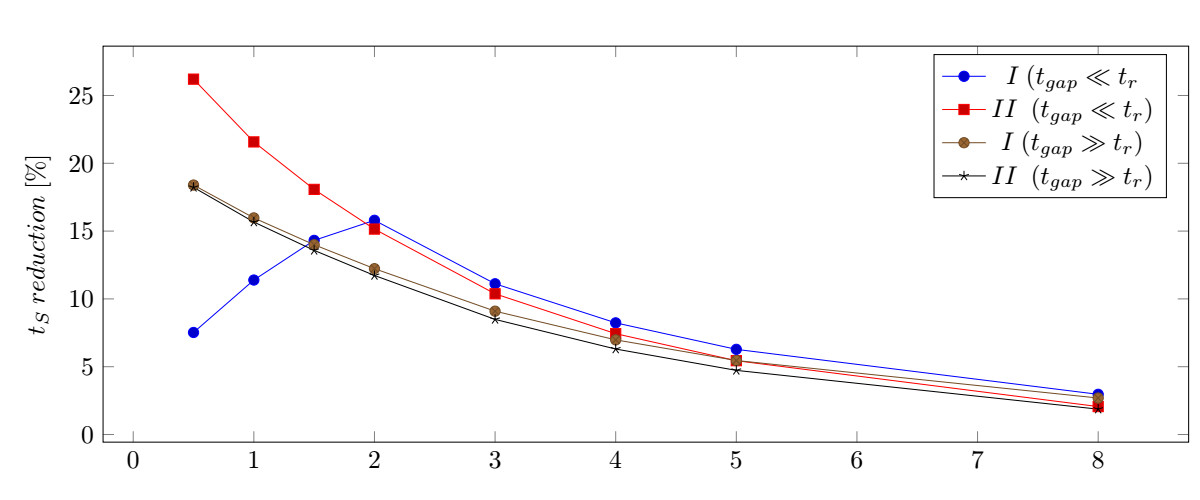}
\end{center}

	\caption{
		Execution-time Improvement vs. Relation-size scaling
		for an average reconfiguration time $t_r$ being drastically smaller or drastically higher
		than the average time gap between the two queries ($t_{gap})$
	}
	\label{fig:execVsScale}
\end{figure}

Fig.~\ref{fig:execVsScale} shows
the improvement in execution time
for each proposed optimization
with varying relation size.
Two different time gaps between the queries are included.

It can be clearly seen that no single optimization is superior.
While II drastically outperforms I for small relation sizes
where $t_{trans} + t_{gap,Q_0} \leq t_{r,acc0}$,
up to a relation size with a scale factor of 2,
I dominates from that point on,
as the reconfiguration time can be hidden completely.

\begin{figure}[htpb]
	\begin{center}
%
%
%
%
 \includegraphics[width=\linewidth, height=.35\columnwidth]{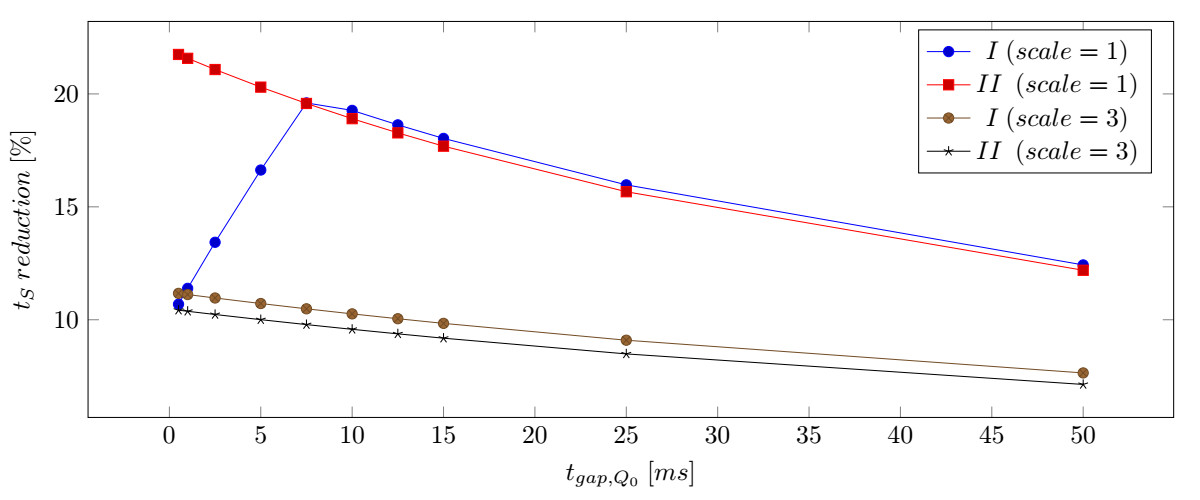}
\end{center}
	\caption{Execution time Improvement vs. $t_{gap}$ for two relation size scales}
	\label{fig:execVstgap}
\end{figure}

A similar behavior can be found in Fig.~\ref{fig:execVstgap}
where $t_{gap,Q_0}$ has been varied.
For small relation sizes and only very little time between the two queries,
optimization II is preferable
and can lead to a runtime improvement of up to 22\%.

While the result sizes vary with the parameter values in the queries,
the time gap $t_{gap,Q_0}$ obtained by the sequence analysis
gives a good indication
of the case where the speculative reconfiguration approach should be used.

\section{Conclusion}


The paper has introduced the utilization of information on query sequences
in the optimization of processing them on reconfigurable accelerator hardware.
The ReProVide Processing Unit (RPU)
is such a system
that can filter data on their way from storage to DBMS.
The reconfiguration required to adapt to the next query
can be substantially reduced
by taking the sequence of coming queries into account.
It can (I) be done in parallel to the result-data transfer of a running query
and can (II) be avoided completely by swapping accelerators.
The second optimization has strong improvement effects for smaller relations,
while the first overtakes for larger relations and/or rising time gaps between queries.

The optimizations done by hand in these evaluations
already show promising results.
So we will be built them into the optimizer.


For further optimizations,
we will use query-analysis graphs similar to those proposed in \cite{Chen98a}.
The idea is to keep result data in the memory of the RPU,
if they can be reused in subsequent queries.
For instance, if one query asks for tuples with $A > 100$ and the next asks for $A > 200$,
the result of the first can be used to generate the result of the second.

More hints can be imagined
to optimize the RPU further:
We could synthesize other accelerators
by combining other arithmetic and comparison operators,
depending on the frequency of these combinations in the query sequences.

\subsubsection*{Acknowledgements}

This work has been supported by the German Science Foundation
(Deutsche Forschungsgemeinschaft, DFG)
with the grant no.\ ME~943/9-1.

\bibliographystyle{unsrt}  

%
%
%
%

\bibliographystyle{ACM-Reference-Format}
\bibliography{references.bib}
\end{document}